\documentclass[%
 reprint,
 amsmath,amssymb,
 aps,
]{revtex4-2}
\usepackage{tikz}
\usepackage{lipsum}
\usepackage{graphicx}
\usepackage{dcolumn}
\usepackage{bm}
\usepackage{subcaption}

\begin{document}

\preprint{APS/123-QED}

\title{Gravitational Lensing Effect from The Revised Deser-Woodard Nonlocal Gravity}

\author{Haida Li}
\email{eqwaplay@scut.edu.cn}
\affiliation{School of Physics and Optoelectronics, South China University of Technology, Guangzhou 510641, China}

\author{Xiangdong Zhang} 
\email{Corresponding author: scxdzhang@scut.edu.cn}
\affiliation{School of Physics and Optoelectronics, South China University of Technology, Guangzhou 510641, China}

\begin{abstract}
    We investigate the gravitational lensing effects of a static spherically symmetric black hole (BH) within the framework of the revised Deser-Woodard (D-W) nonlocal gravity. By analyzing the deflection angle in both the weak and strong field limits, we derive several distinguishing features of the model. In the weak field limit, we report a leading-order correction to the deflection angle directly attributed to the non-local nature of the theory. In the strong field limit, we find that the lensing corrections are almost linearly dependent on the coupling parameter $\zeta$ while being exponentially suppressed by the exponent parameter $n$. Furthermore, the gravitational lensing effect in the revised D-W model at a given time shares similar scale-invariant behavior to General Relativity and conformal gravity, offering a potential pathway to distinguish it from other alternatives using astronomical observations.
\end{abstract}

\maketitle


\section{\label{sec:111}Introduction}

As one of the most profound puzzles in modern physics, the discovery of the late-time accelerated expansion of the Universe \cite{SupernovaCosmologyProject:1998vns,SupernovaSearchTeam:1998fmf,Peebles:2002gy, Frieman:2008sn,DAgostino:2019wko} signals a potential breakdown of General Relativity on cosmological scales. The standard $\Lambda$CDM model provides a phenomenologically successful description of this acceleration through the inclusion of a cosmological constant. However, it suffers from severe theoretical issues, most notably the fine-tuning problem \cite{Weinberg:1988cp,Padmanabhan:2002ji,DAgostino:2022fcx}, i.e., the observed value of $\Lambda$ diverges from quantum field theory predictions by over 100 orders of magnitude, as well as the cosmic coincidence problem \cite{Zlatev:1998tr,Garriga:2002tq,Velten:2014nra}. These fundamental inconsistencies have catalyzed the search for alternative gravitational theories.

Deser-Woodard (D-W) gravity \cite{Deser:2007jk} is a non-local modification of General Relativity proposed to naturally explain the late-time acceleration of the Universe through infrared quantum corrections, essentially replacing the cosmological constant with a non-local distortion function $f(\square^{-1}R)$ in the Einstein-Hilbert action. This theory avoids the fine-tuning problems of the standard $\Lambda$CDM model, with the function $f$ phenomenologically reconstructed to reproduce the observed expansion history. A revised version of the D-W model is later introduced \cite{Deser:2019lmm}. In the revised model, the nonlocal function is constrained to reproduce the observed cosmic evolution while remaining compatible with solar system tests. Recently, the static spherically symmetric solution of the revised D-W model was obtained \cite{DAgostino:2025wgl} by inserting an ansatz on the leading order $\frac{1}{r^n}$ correction to the time component of the metric tensor. Also, there have been several works studying the gravitational effects of the obtained static spherically symmetric solution \cite{DAgostino:2025sta,Neves:2025uoi,Liu:2025cpp,DAgostino:2026wln}.

The BH gravitational lensing \cite{Virbhadra:1999nm,Bozza:2002zj,Virbhadra:2008ws,Tsukamoto:2016qro,Fu:2021fxn,Igata:2025glk} can potentially serve as a testing ground for different models of modified gravity. Beyond the conventional photon deflections around the BH, there are many works studying the effects of time-like particle trajectories deflected by the BH. For example, the gravitational lensing effect of massive particles with non-zero spin has already been investigated in \cite{Zhang:2022rnn,Li:2025mcp}. Meanwhile, the particle motion around a Kerr black hole has also been investigated previously, although not focusing on the critical radius, by some works \cite{Zhang:2018ocv,Liu:2019wvp,Wu:2021pgf}. As a direct astronomically observable quantity, the critical radius has been theorized as a tool for testing the effect of quantum gravity, e.g., in \cite{Li:2024afr,Li:2025zdt}. 

In this work, by computing the gravitational lensing effects in both the weak and strong field limit, we aim to expand the knowledge on the corrections induced by the non-local terms in the revised D-W model. There are several key results we achieve: First, we discover that the revised D-W model contains static spherically symmetric BH solutions that enable the leading order corrections to weak gravitational lensing. Second, for the strong gravitational lensing, we show that the correction depends almost linearly on the coupling parameter $\zeta$, while being suppressed exponentially as $n$ increases. This fact suggests that it is crucial to determine the actual order of $n$ in the revised D-W gravity. Also, for observables, we discover that the correction of $\theta$ is the largest among all three observables. And, as $\zeta\rightarrow 0$, the corrections of $s$ and $\theta$ overlap. Third, we also discuss the classification of two types of modified gravity theories by examining their behavior in terms of scale invariance of the gravitational lensing effect: the ones that share the same scale-invariant behavior as general relativity and the ones that have different scale-invariant behavior from general relativity. We find that the revised D-W theory, due to its coupling parameter $\zeta$ being dimensionless in the spatial directions, belongs in the first category.

The structure of this work is as follows. In Section \ref{sec2}, we briefly introduce the revised D-W gravity and the most recent static spherical symmetric solutions obtained by inserting a specific ansatz. In Section \ref{sec3}, we derive the weak field limit of the gravitational lensing effect of a spherically symmetric BH in the revised D-W model, where unusual leading order corrections related to the non-local feature of the revised D-W model are discovered. In Section \ref{sec4}, several key results regarding the strong field limit will be presented, including the accuracy of both the weak and strong field limit, the corrections to the lensing image, the deflection angle, as well as the corrections to three key observables of gravitational lensing that can be directly extracted from the strong field limit. Finally, we will provide some discussion on the results we obtained in this work in Section \ref{sec5}. In this work, we use the geometric unit such that $c=\hbar=G=1$ and set black hole mass $M=1$.

\section{BLACK HOLE SOLUTION IN NONLOCAL GRAVITY}\label{sec2}

The action of the revised D-W gravity can be written as \cite{Deser:2019lmm}:
\begin{equation}\label{action1}
S=\frac{1}{16 \pi} \int \mathrm{~d}^4 x \sqrt{-g} R(1+F[Y]),
\end{equation}
where $R$ is the Ricci scalar, and $g$ denotes the determinant of the metric tensor, $g_{\mu\nu}$. The non-local modification is generated by considering the following auxiliary fields:
\begin{equation}
    \begin{aligned}
\square X & =R, \\
\square Y & =g^{\mu \nu} \partial_\mu X \partial_\nu X, \\
\square U & =-2 \nabla_\mu\left(V \nabla^\mu X\right), \\
\square V & =R f_{, Y},
\end{aligned}
\end{equation}
where $\square=g^{\mu \nu} \nabla_\mu \nabla_\nu$ is the d’Alembert operator such that:
\begin{equation}
\square u \equiv \frac{1}{\sqrt{-g}} \partial_\alpha\left[\sqrt{-g} \partial^\alpha u\right]
\end{equation}
The equations of motion of the theory can be obtained by varying the action (\ref{action1}) with respect to the metric \cite{DAgostino:2025wgl}:
\begin{equation}
\left(G_{\mu \nu}+g_{\mu \nu} \square-\nabla_\mu \nabla_\nu\right) W+\mathcal{K}_{(\mu \nu)}-\frac{1}{2} g_{\mu \nu} g^{\alpha \beta} \mathcal{K}_{\alpha \beta}=0,
\end{equation}
where $G_{\mu \nu}=R_{\mu \nu}-\frac{1}{2} g_{\mu \nu} R$, $W=1+U+F[Y]$, and:
\begin{equation}
\mathcal{K}_{\mu \nu}=\partial_\mu X \partial_\nu U+\partial_\mu Y \partial_\nu V+V \partial_\mu X \partial_\nu X,
\end{equation}
and we denote its symmetrization as $\mathcal{K}_{(\mu \nu)}=\frac{1}{2}\left(\mathcal{K}_{\mu \nu}+\mathcal{K}_{\nu \mu}\right)$. 

Static spherically symmetric solutions to the model have the following general form \cite{DAgostino:2025wgl} :
\begin{equation}\label{metricS}
d s^2=-A(r) d t^2+B(r) d r^2+C(r)\left(d \theta^2+\sin ^2 \theta d \phi^2\right).
\end{equation}
By introducing the ansatz: $A(r) =1-\frac{2}{r}-\frac{\zeta}{r^n}$, one may then obtain the solution up to first order corrections of $\zeta$ as:
\begin{equation}\label{solutionG}
\begin{aligned}
A(r) & =1-\frac{2}{r}-\frac{\zeta}{r^n}, \\
(B(r))^{-1} & =1-\frac{2}{r}+\frac{\zeta }{3^n r^{n+1}\left(r-3\right)^2}\\
\times & \left\{3^n r\left[n\left(r-3\right)\left(r-2\right)+4 r-9\right]\right. \\
&\left.\quad-3\left(r-2\right)\left(2r-3\right)\left(r\right)^n\right\}, \\
C(r) & =r^2 .
\end{aligned}
\end{equation}
This family of solutions satisfies the condition of asymptotic flatness and is free from essential singularities outside the event horizon. Also, it is important to note that since the parameter $\zeta$ represents cosmological expansion. For any given cosmological epoch, it is independent of the local length scale. Therefore, the gravitational lensing effect for photons generated by the revised D-W black hole at a given time is scale-invariant, while the gravitational lensing effect for massive particles can still be scale-variant due to the coupling between the mass of the test particle and the specific size of the central black hole. This behavior is similar to general relativity, $f(R)$ theory \cite{Buchdahl:1970ldb} and conformal gravity \cite{weyl1918gravitation,Bach:1921zdq,mannheim1989exact}, but is different from the corrections of lensing effect coming from, e.g., the properties of the BH such as charge \cite{Jusufi:2015laa,Bjerrum-Bohr:2016hpa} or spin, as well as BH models inspired by theories of quantum gravity \cite{Fu:2021fxn,Li:2025zdt}, in which the quantum parameter is usually associated with the Planck length. 

\section{Weak Field Limit}\label{sec3}

In a general static spherically symmetric spacetime with the metric given in eqn. (\ref{metricS}), the light-like geodesics in the equatorial plane ($\theta=\frac{\pi}{2}$) is given by:
\begin{equation}
-A(r) \dot{t}^2+B(r) \dot{r}^2+C(r) \dot{\phi}^2=0 .
\end{equation}
The impact parameter of the test particle (photon) can be determined by the following relation with respect to the nearest distance $r_0$ of the photon to the center black hole in its entire trajectory:
\begin{equation}\label{impact}
    b(r_0)=\sqrt{\frac{C(r_0)}{A(r_0)}},
\end{equation}
The deflection angle $\alpha(r_0)$ can then be computed as:
\begin{equation}
\alpha\left(r_0\right)=I\left(r_0\right)-\pi,
\end{equation}
where:
\begin{equation}\label{dangle}
I\left(r_0\right) \equiv 2 \int_{r_0}^{\infty} \frac{d r}{\sqrt{\frac{C}{B}\left(\frac{C}{A b^2}-1\right)}}.
\end{equation}
For most theories, this integration can be easily evaluated numerically. However, we would like to obtain the analytical asymptotic behaviors in both the strong field limit and the weak field limit. This is because both limits are well established theoretically for spherically symmetric spacetime, and the coefficients extracted from these two limits can be directly associated with physical observables that can be directly compared with astronomical observations.

For the weak field limit, namely for $1/r_0\rightarrow 0$, we can introduce the following change of variable \cite{Fu:2021fxn}: $z=\frac{r_0}{r}$, then the integration becomes:
\begin{equation}
I\left(r_0\right) = \int_{0}^{1} d z \left(\frac{2 r_0}{z^2}/\sqrt{\frac{C(z)}{B(z)}\left(\frac{C(z)}{A(z) b^2}-1\right)}\right).
\end{equation}
Using eqn. (\ref{impact}), the integration can be computed order by order after expanding the integrand in terms of $\frac{1}{r_0}$ to obtain:
\begin{equation}
\begin{split}
    \alpha(r_0)&=(4 + \frac{2 \zeta}{3}) \frac{1}{r_0} -(4-\frac{15\pi}{4} + \frac{(8-15\pi) a1}{12})\frac{1}{r_0^2}\\
    &+  (\frac{122}{3}-\frac{15\pi}{2} + \frac{38-5\pi \zeta}{2} )\frac{1}{r_0^3}+\mathcal{O}(\frac{1}{r_0^4})
\end{split}
\end{equation}

Moreover, after expanding $b$ in terms of $\frac{1}{r_0}$, we can also obtain its inversed series:
\begin{equation}
    \begin{split}
        \frac{1}{r_0}=f(\frac{1}{b})
    \end{split}
\end{equation}
Finally, we can express the weak deflection angle as a function of the impact factor $b$. Here we provide the results for several cases of $n$ (up to the first order corrections in $\zeta$):
\begin{widetext}
\begin{equation}\label{ncorrect}
    \begin{split}
        n&=2:\quad \alpha=(4+\frac{2 \zeta}{3})\frac{1}{b}+(\frac{15 \pi }{4}+\frac{5 \pi  \zeta}{4})\frac{1}{b^2}+(\frac{128}{3}+\frac{64
   \zeta}{3})\frac{1}{b^3}+\mathcal{O}(\frac{1}{b^4})\\
           n&=3:\quad \alpha=(4+\frac{2 \zeta}{9})\frac{1}{b}+(\frac{15 \pi }{4}+\frac{5 \pi  \zeta}{12})\frac{1}{b^2}+(\frac{128}{3}+\frac{64
   \zeta}{9})\frac{1}{b^3}+\mathcal{O}(\frac{1}{b^4})\\
           n&=4:\quad \alpha=(4+\frac{2 \zeta}{27})\frac{1}{b}+(\frac{15 \pi }{4}+\frac{5 \pi  \zeta}{36})\frac{1}{b^2}+(\frac{128}{3}+\frac{64
   \zeta}{27})\frac{1}{b^3}+\mathcal{O}(\frac{1}{b^4})\\
           n&=10:\quad \alpha=(4+\frac{2 \zeta}{19683})\frac{1}{b}+(\frac{15 \pi }{4}+\frac{5 \pi  \zeta}{26244})\frac{1}{b^2}+(\frac{128}{3}+\frac{64
   \zeta}{19683})\frac{1}{b^3}+\mathcal{O}(\frac{1}{b^4})\\
    \end{split}
\end{equation}
\end{widetext}
There are two main facts that can be extracted from eqn. (\ref{ncorrect}). Firstly, $\frac{1}{b}$-order corrections are present for all of the $n$ values we studied. This correction is at lower order (larger) when comparing with the corrections coming from either quantum corrected black holes \cite{Fu:2021fxn,Li:2025zdt} or properties of black holes in GR, e.g., the charge or spin of a black hole \cite{Jusufi:2015laa}. This fact is naturally expected because of the non-local feature of the revised D-W model. Since the D-W model modifies general relativity by matching quantum loop corrections with the current rate of cosmic acceleration. Therefore, in this model, the effect of cosmic acceleration can be seen locally in weak gravitational lensing. Secondly, it can be seen from eqn. (\ref{ncorrect}) that the correction decreases as the parameter $n$ becomes larger. Interestingly, the corrections of each order in $\frac{1}{b}$ (for any value of $n$) are exactly proportional to their corresponding corrections of the $n=2$ case by a universal factor of $3^{(n-2)}$ across every order of $\frac{1}{b}$. This unique factor results directly from the overall $3^n$ in the denominator of $(B(r))^{-1}$.  

\section{Strong Field Limit}\label{sec4}

Now we compute the strong BH gravitational lensing effect from the revised D-W model. Let the closest distance to the lens of a photon trajectory be $r_0$, the minimum such distance is the BH photon sphere $r_m$ satisfying \cite{Virbhadra:1999nm}:
\begin{equation}\label{rps}
\frac{C^{\prime}(r_m)}{C(r_m)}=\frac{A^{\prime}(r_m)}{A(r_m)}.
\end{equation}
For $r_0\rightarrow r_m$, the gravitational lensing deflection angle (\ref{dangle}) can be approximated by the strong field limit. A general approach to obtain the strong field limit for a BH whose line elements take the form (\ref{metricS}) is described in \cite{Bozza:2002zj}. As an alternative to the original proposal, we consider the variable $z=1-\frac{r_0}{r}$ used in \cite{Fu:2021fxn,Liu:2024wal}, which allows for the direct conversion between $z$ and $r$. The integral can then be rewritten as:
\begin{equation}
\begin{aligned}
I\left(r_0\right) & =\int_0^1 R\left(z, r_0\right) f\left(z, r_0\right) d z, \\
\end{aligned}
\end{equation}
where:
\begin{equation}
\begin{aligned}
R\left(z, r_0\right) & =\frac{2 r^2 \sqrt{A(r) B(r) C_0}}{r_0 C(r)} \\
f\left(z, r_0\right) & =\frac{1}{\sqrt{A_0-\frac{A(r) C_0}{C(r)}}},
\end{aligned}
\end{equation}
where $A_0:=A(r_0)$ and $C_0:=C(r_0)$. $R(z,r_0)$ is regular for  $r\geq r_0\geq r_m$, while $f(z,r_0)$ is divergent in $z$ as $z\rightarrow 0$, i.e., $r\rightarrow r_0$. The divergent term $f(z,r_0)$ can be expanded up to second order of $z$ as:
\begin{equation}
f\left(z, r_0\right) \sim f_0\left(z, r_0\right)=\frac{1}{\sqrt{c_1\left(r_0\right) z+c_2\left(r_0\right) z^2}},
\end{equation}
where:
\begin{widetext}
\begin{equation}
\begin{aligned}
c_1\left(r_0\right)=-r_0 A_0^{\prime}+\frac{r_0 A_0 C_0^{\prime}}{C_0},\ c_2\left(r_0\right)=r_0 \frac{-2 r_0 A_0 C_0^{\prime 2}-C_0^2\left(2 A_0^{\prime}+r_0 A_0^{\prime \prime}\right)+C_0\left[2 r_0 A_0^{\prime} C_0^{\prime}+A_0\left(2 C_0^{\prime}+r_0 2 C_0^{\prime \prime}\right)\right]}{2 C_0^2}
\end{aligned}
\end{equation}
\end{widetext}
which captures the dominant contribution of the divergence as $z\rightarrow 0$.

Using this expansion, the integral $I(r_0)$ can be split into regular part $I_R(r_0)$ and divergent part $I_D(r_0)$ as:
\begin{equation}
\begin{aligned}
I\left(r_0\right) & =I_D\left(r_0\right)+I_R\left(r_0\right) \\
I_D\left(r_0\right) & =\int_0^1 R\left(0, r_{m}\right) f_0\left(z, r_0\right) d z \\
I_R\left(r_0\right) & =\int_0^1\left[R\left(z, r_0\right) f\left(z, r_0\right)-R\left(0, r_{m}\right) f_0\left(z, r_0\right)\right] d z.
\end{aligned}
\end{equation}
As a result, the deflection angle can be computed in terms of the impact parameter $b$ as:
\begin{equation}\label{stronga}
\alpha(b)=-a \ln \left(\frac{b}{b_m}-1\right)+u+\mathcal{O}\left[\left(b-b_m\right) \ln \left(b-b_m\right)\right],
\end{equation}
where:
\begin{equation}
\begin{aligned}
a& \equiv \frac{R\left(0, r_m\right)}{2 \sqrt{c_2\left(r_m\right)}}, \\
u & \equiv a \ln \delta+I_R\left(r_0\right)-\pi, \\
\delta & \equiv r_m^2\left(\frac{C^{\prime \prime}\left(r_m\right)}{C\left(r_m\right)}-\frac{A^{\prime \prime}\left(r_m\right)}{A\left(r_m\right)}\right).
\end{aligned}
\end{equation}
Despite the dependence of the impact factor $b$ in $\alpha(b)$, using the factors $a$ and $u$, several $b$-independent lens observables can also be extracted \cite{Bozza:2002zj,Fu:2021fxn,Liu:2024wal}:
\begin{itemize}
\item Angle $\theta_{\infty}$ of the innermost image (photon ring). This angle is defined as the observed angle between the center BH and the innermost image of the source. Since the photon trajectory can go around the lens multiple times before finally reaching the observer, there can be a total of $n$ images of the source. $n$ is not bounded since the deflection angle $\theta_n$ is unbounded for $b\rightarrow b_m$, namely when $r_0$ approximate the BH's photon ring. A limit $\theta_{\infty}\approx \frac{b_m}{D_{OL}}$ can be obtained by taking $n\rightarrow \infty$.
\item The angular separation $s$ between the outermost image and the innermost image (the lower bound of the series of images as $n\rightarrow$, since these images are unlikely to be distinguishable):
\begin{equation}
s := \theta_1-\theta_{\infty} \sim \theta_{\infty} \mathrm{e}^{\frac{u-2 \pi}{a}}.
\end{equation}

\item the quotient $\mu$ of the flux of the outermost relativistic image to that of all other relativistic images:
\begin{equation}
\mu \sim \frac{\left(\mathrm{e}^{\frac{4 \pi}{a}}-1\right)\left(\mathrm{e}^{\frac{2 \pi}{a}}+\mathrm{e}^{\frac{u}{a}}\right)}{\mathrm{e}^{\frac{4 \pi}{a}}+\mathrm{e}^{\frac{2 \pi}{a}}+\mathrm{e}^{\frac{u}{a}}}
\end{equation}

\end{itemize}
These $b$-independent lens observables are only dependent on the properties of the spacetime, thus serving as an ideal testing ground for studying the corrections of models of modified gravity. 

\begin{figure}[h!]
\includegraphics[height=6cm]{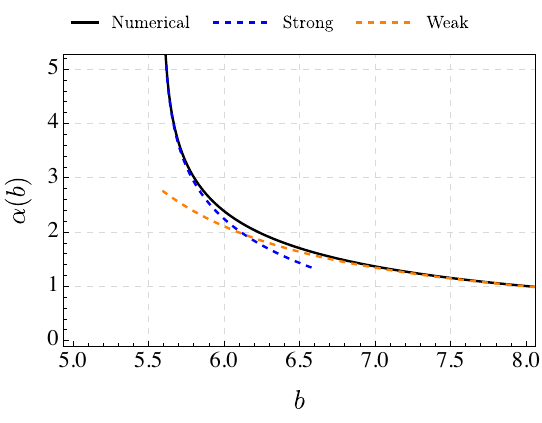} 
\caption{\label{fig:1} Comparison between results obtained via different methods ($n=2,\ \zeta=0.5$): The numerical results obtained by directly performing the integration (\ref{dangle}) (Black line). Strong field limit (Blue dashed line). Weak field limit (Orange dashed line).}
\end{figure}

Fig. \ref{fig:1} shows the deflection angle $\alpha$ with respect to the impact factor $b$ for the case of $n=2$, $\zeta=0.5$. In this graph, we show in blue and orange dashed lines that both the weak field limit in (\ref{ncorrect}) and the strong field limit in (\ref{stronga}) converge to the actual deflection angle (Black line) when $b\rightarrow\infty$ and $b\rightarrow b_m$, respectively. 

\begin{figure*}[h!tb]
    \begin{subfigure}[b]{0.4\textwidth}
        \centering
        \includegraphics[height=6cm]{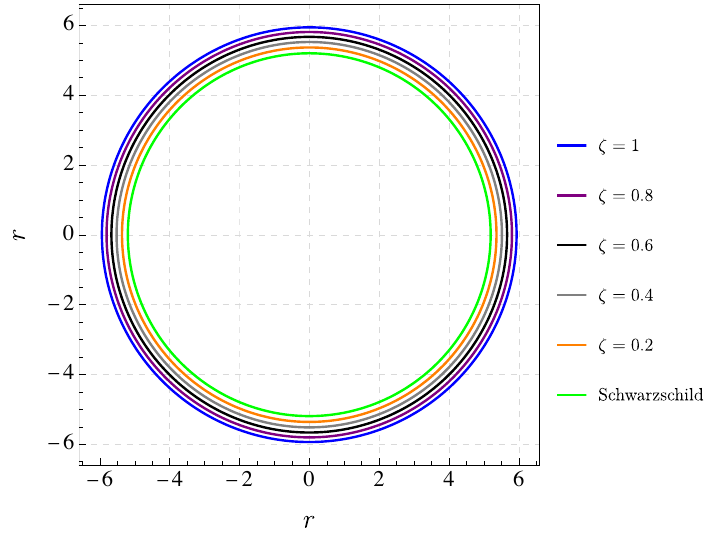} 
        \caption{}
        \label{fig:sub1}
    \end{subfigure}
    \hspace{+1cm}
    \begin{subfigure}[b]{0.4\textwidth}
        \centering
        \includegraphics[height=5.5cm]{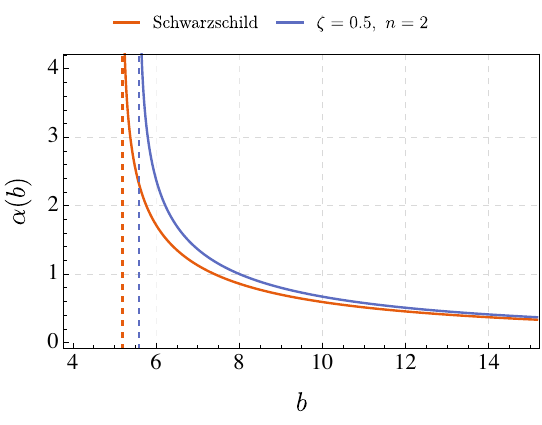} 
        \caption{}
        \label{fig:sub2}
    \end{subfigure}
    \hspace{+1cm}
    \begin{subfigure}[b]{0.4\textwidth}
        \centering
        \includegraphics[height=5cm]{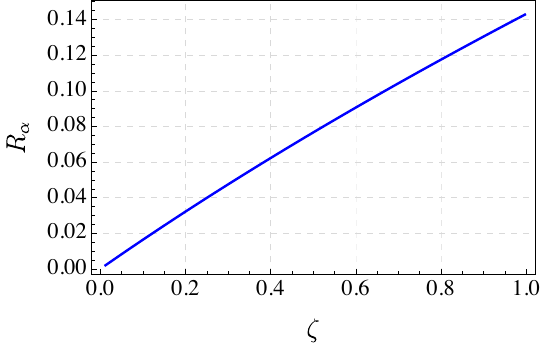} 
        \caption{}
        \label{fig:sub3}
    \end{subfigure}
    \hspace{+1cm}
    \begin{subfigure}[b]{0.4\textwidth}
        \centering
        \includegraphics[height=5cm]{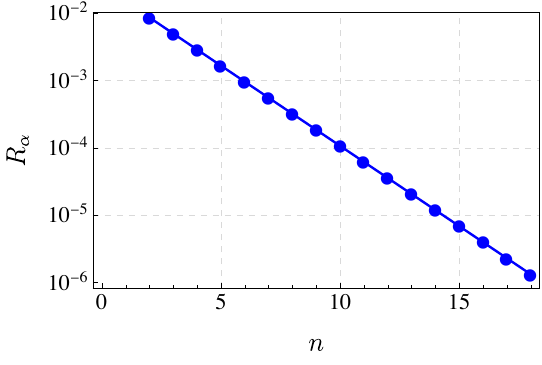} 
        \caption{}
        \label{fig:sub4}
    \end{subfigure}
    \hspace{+1cm}
    \begin{subfigure}[b]{0.4\textwidth}
        \centering
        \includegraphics[height=5cm]{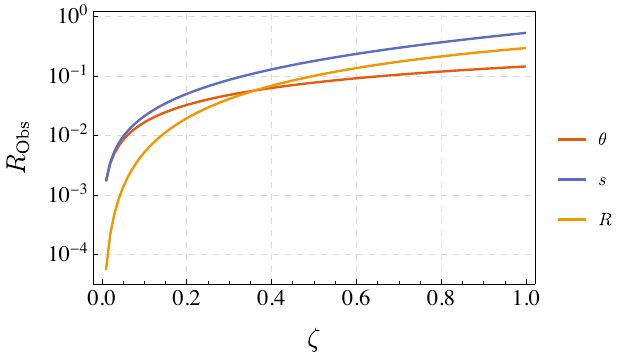} 
        \caption{}
        \label{fig:sub5}
    \end{subfigure}
    \hspace{+1cm}
    \begin{subfigure}[b]{0.4\textwidth}
        \centering
        \includegraphics[height=5cm]{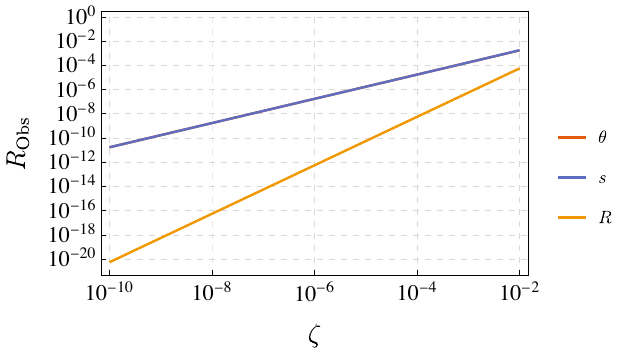} 
        \caption{}
        \label{fig:sub6}
    \end{subfigure}
\caption{The main results of our work: (a) The photon rings of models of different parameters, with the innermost one being the Schwarzschild BH. (b) Corrections to the strong field limit for $\zeta=0.5,\ n=2$ compared to the Schwarzschild BH. (c) Relations between the relative difference $R_{\alpha}$ of deflection angle and the coupling parameter $\zeta$, with $n=2$ (d) Relations between the relative difference $R_{\alpha}$ of deflection angle and $n$, with $\zeta=1$. (e) Relations between the relative difference $R_{\mathrm{Obs}}$ of observables and the coupling parameter $\zeta$, with $n=2$. (f) Log-Log Continuation of (e) to the regime where $\zeta$ is extremely small. }
\label{fig:2}
\end{figure*}

In Fig. \ref{fig:sub1}, the photon ring of the BHs of several different parameters is shown by taking $n=2$ and $\zeta=0.2,\ 0.4,\ 0.6,\ 0.8,\ 1.0$. From this graph, we can see that as the coupling parameter $\zeta$ becomes larger, the radius of the photon ring becomes increasingly larger. Interestingly, since the radius of the BH horizon is also increasing, the overall effect for positive $\zeta$ is actually slightly weakening the deflection. This can be easily seen by looking at the data when $\zeta=0.5,\ n=2$, where $r_H=1.11237 r_{H,\mathrm{Sch}}$, while $r_m=1.10093 r_{m,\mathrm{Sch}}$, where "Sch" denotes the results for the Schwarzschild BH.

Fig. \ref{fig:sub2} shows the deflection angle $\alpha$ of both the static spherically symmetric revised DW black hole model ($n=2$ and $\zeta=0.5$) and Schwarzschild BH w.r.t. the impact parameter $b$. Similar to Fig. \ref{fig:sub1}, the same effect of increased defection angle for the revised DW model can be observed as $b\rightarrow b_m$, where the nearest distance $r_0$ to the center of the black hole during the photon's entire trajectory approaches the black hole's photon ring.

Also in Fig. \ref{fig:sub2}, it is obvious that the difference is larger in the strong field limit than in the weak field limit, despite the leading order corrections appearing in the weak field limit. From eqn. (\ref{ncorrect}) and Fig. \ref{fig:1}, it is also shown that in the weak field limit, there is a fixed correction at every order of $\frac{1}{b}$ for each value of $\zeta$ and $n$.

Moreover, for the case of the deflection angle, we consider the case where the impact factor is very near to the innermost possible impact factor $b_m=5.19615$ for each BH. A $\mathrm{Log}_{10}$-$\mathrm{Log}_{10}$ graph (Fig. \ref{fig2}) can be produced in this region by defining the relative difference $R_\alpha$ with respect to the impact factor $b$:
\begin{equation}
    R_\alpha:=\frac{|\alpha-\alpha_{Sch}|}{\alpha_{Sch}}.
\end{equation}
Where $\alpha_{Sch}$ denotes the deflection angle with the exact same impact factor $b$ in the Schwarzschild case.

Next, we plot the trend of $R_\alpha$ with respect to $\zeta$ and $n$ in Fig. \ref{fig:sub3} and Fig. \ref{fig:sub4}, respectively. In Fig. \ref{fig:sub3}, it can be seen that $R_\alpha$ decreases almost linearly w.r.t $\zeta$ as $\zeta\rightarrow 0$. On the other hand, in Fig. \ref{fig:sub3} we show that the relative correction $R_\alpha$ decreases much faster in $n$ than in $\zeta$. This fact suggests that when considering the real physical implications of revised D-W non-local gravity, fixing the leading order of $r_n$ is extremely important in determining its deviation from Schwarzschild spacetime.

\begin{figure*}[h!tb]
    \centering
    \begin{subfigure}[b]{0.4\textwidth}
        \centering
        \includegraphics[height=6cm]{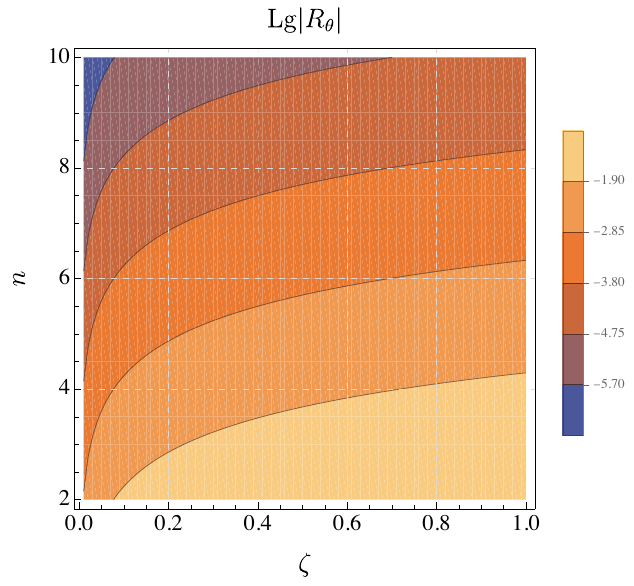} 
        \caption{}
        \label{fig:sub7}
    \end{subfigure}
    \hspace{+0.5cm}
    \begin{subfigure}[b]{0.4\textwidth}
        \centering
        \includegraphics[height=6cm]{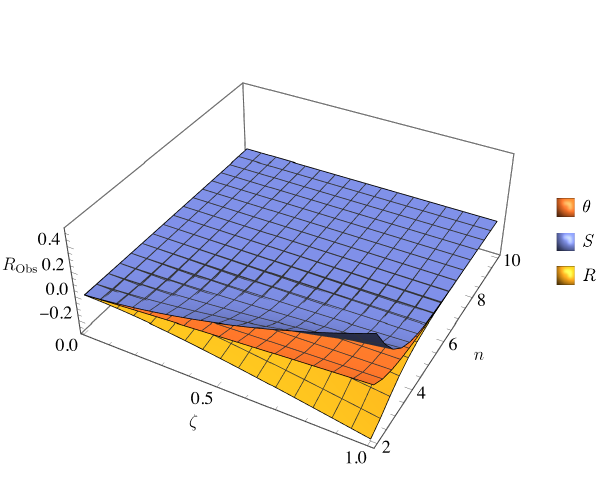} 
        \caption{}
        \label{fig:sub8}
    \end{subfigure}
\caption{(a) The contour plot of the dependence of $|R_{\theta}|$ on both $\zeta$ and $n$. (b) The 3D plot of the dependence of $R_{obs}$ for all three observables on both $\zeta$ and $n$.}
\label{fig:3}
\end{figure*}

For the observables, we can also define the relative difference from the Schwarzschild BH:
\begin{equation}
    R_{\mathrm{Obs}}:=\frac{|\mathrm{Obs}-\mathrm{Obs}_{Sch}|}{\mathrm{Obs}_{Sch}}.
\end{equation}
with $\mathrm{Obs}=\theta, S, R$ and $\mathrm{Obs}_{Sch}$ is the observables for Schwarzschild BH. The results for observables are shown in Fig. \ref{fig:sub5}, Fig. \ref{fig:sub6}, Fig. \ref{fig:sub7}, and Fig. \ref{fig:sub8}.

In Fig. \ref{fig:sub5} and Fig. \ref{fig:sub6}, we show the relative difference $R_{\mathrm{Obs}}$ of all three observables with respect to different coupling parameter $\zeta$ ($n$ is set to be 2 for all cases). In Fig. \ref{fig:sub5}, we plot the results for $0.01\leq\zeta\leq 1$. As we can see from this graph, when $\zeta$ is relatively large, the correction of $R$ is larger than the correction of $s$, while when $\zeta$ gets smaller, the correction on $R$ becomes the smallest in all three observables. Also, throughout the range $0.01\leq\zeta\leq 1$, the correction of $\theta$ is the largest among all three observables. However, we are beginning to see that this correction is getting closer to the correction of $s$. They almost overlap as $\zeta\rightarrow 0$.

In Fig. \ref{fig:sub6}, we continue to investigate the trend using a Log-Log plot when $\zeta$ gets very small. As is verified in this graph, the corrections of $\theta$ and $s$ indeed overlap as the coupling parameter becomes very small. In the meantime, the correction of $\theta$ gets even smaller, making its detection much less possible than the corrections of the other two observables.

In Fig. \ref{fig:sub7} and \ref{fig:sub7}, we plot the contour plot of $\mathrm{Lg}|R_\theta|$ for different combinations of $n$ and $\zeta$. Similar to the results we have shown so far, the largest $\mathrm{Lg}|R_\theta|$ appears when $n=2$ and $\zeta=1$. In particular, $\mathrm{Lg}|R_\theta|$ is suppressed linearly with respect to $\zeta$ and exponentially with respect to $n$. Fig. \ref{fig:sub8} shows the corrections of all three types of observables. It can be seen here that both $\theta$ and $s$ are positively impacted by larger values of $\zeta$ and smaller values of $n$, while $R$ is negatively impacted.

Finally, we should also note that instead of using any specific settings, e.g., solar mass BH, Saggitarius A, or M87, we only specified that $D_{OL} = 10^10 r_H$, without fixing $r_H$. This is because the fact that, similar to the Schwarzschild black hole, the revised D-W model is also scale invariant for the gravitational lensing effect of photons. Again, for many corrections which may be larger or at similar levels as the correction coming from non-local gravitational effect, the gravitational lensing effect of photon is not scale invariant for different BH masses because of the dimension of the coupling parameters, e.g., charge, magnetic field, corrections from quantum geometry, etc.

\section{Discussion}\label{sec5}
In this work, we have studied the gravitational lensing effect from the revised D-W non-local gravity. Using both the weak field limit and the strong field limit method, we successfully computed the relation between the total deflection angle and the initial impact parameter for different parameters of a static spherically symmetric BH in the revised D-W model. In the meantime, we have also computed the lens observables from these models. Summarizing all of the results, we have made the following three key discoveries:

First, we discover that the revised D-W model contains static spherically symmetric BH solutions that enable the leading order corrections to weak gravitational lensing. Additionally, we have also investigated the reason for the $3^n$-proportional corrections across all higher orders of $\frac{1}{b}$. It turns out that the $3^n$ term comes from the overall $3^n$ in the denominator of the correction term in $(B(r))^{-1}$ in eqn.(\ref{solutionG}). This is mainly due to the fact that the leading order term in $ r$ in the numerator contains only a fixed overall coefficient that does not counter the overall $3^n$ factor as $n$ increases.

Second, for the strong gravitational lensing, we begin by showing that, when setting $\zeta$ to be positive, the correction of the non-local term on the lensing effect slightly weakens the deflection. The effect is most noticeable when the impact parameter of the photon is close to the minimum possible impact parameter $b_m$. For the corrections on the deflection angle, it is discovered that the correction depends almost linearly on the coupling parameter $\zeta$ and is suppressed exponentially as $n$ increases, suggesting that it is crucial to determine the actual order of $n$ through first principle in the revised D-W gravity. Also, for observables, we discover that, when $\zeta$ is relatively large, the correction of $R$ is larger than the correction of $s$. As $\zeta$ gets smaller, the correction on $R$ becomes the smallest in all three observables. Throughout the range $0.01\leq\zeta\leq 1$, the correction of $\theta$ is the largest among all three observables. But, as $\zeta\rightarrow 0$, the corrections of $s$ and $\theta$ becomes the same.

Third, note again that in pure general relativity, scale invariance applies to the lensing effect of photons, namely, the entire calculation can be done by setting the BH mass to 1, and the result holds for all BHs. However, the lensing of other massive particles will not share the same scale invariance since the involvement of non-zero test particle mass, spin, etc. Thus, by examining the scale invariance of the trajectories of different types of particles, we can actually classify modified theories of gravity into two different categories: the ones that share the same scale-invariant behavior as general relativity and the ones that have different scale-invariant behavior from general relativity. The revised D-W theory, due to its coupling parameter $\zeta$ being dimensionless in the spatial directions, belongs in the first category.

Finally, it should be further noted that the current study of the revised D-W model remains limited. First, only the static spherically symmetric spacetime admitting a specific ansatz is studied. While such a solution is permissible in the model by considering the quasi-static approximation of a fixed moment in time, it is certainly more interesting to see how the spherically symmetric BH will interact dynamically with the cosmological expansion. Second, although the leading order non-local correction in $r$ should contribute most significantly to the modifying effect, it remains to be seen whether other higher order terms can contribute simultaneously in the same model, and whether a more general solution without admitting any ansatz can generate completely different corrections.

\section*{Acknowledgements}
This work is supported by the National Natural Science Foundation of China (NSFC) with Grant No.12275087.

\providecommand{\noopsort}[1]{}\providecommand{\singleletter}[1]{#1}%

\end{document}